\documentclass[preprint]{aastex}
\usepackage{eucal}
\shortauthors{Watson \& Wiebe} 
\shorttitle{Circular polarization of Maser Radiation from a Turbulent Disk}

\begin{document}

\title{Spectra of Maser Radiation from a Turbulent, Circumnuclear Accretion Disk. III. Circular polarization}

\author{W. D. Watson and D. S. Wiebe\altaffilmark{1}}

\affil{Department of Physics, University of Illinois, 1110 West Green
Street, Urbana, IL 61801-3080}
\email{w-watson@uiuc.edu}
\altaffiltext{1}{Permanent address: Institute of Astronomy of the Russian
Academy of Sciences, 48, Pyatnitskaya str., 109017 Moscow, Russia}

\begin{abstract} 
Calculations are performed for the circular polarization of maser 
radiation from a turbulent, Keplerian disk that is intended to 
represent the sub-parsec disk at the nucleus of the galaxy NGC4258. 
The polarization in the calculations is a result of the Zeeman 
effect in the regime in which the Zeeman splitting is much less 
than the spectral linebreadth. Plausible configurations for 
turbulent magnetic and velocity fields in the disk are created 
by statistical methods. This turbulence, along with the Keplerian 
velocity gradients and the blending of the three hyperfine components 
to form the $6_{16} - 5_{23}$ masing transition of water, are key ingredients 
in determining the appearance of the polarized spectra that are 
calculated. These spectra are quite different from the polarized 
spectra that would be expected for a two-level transition where 
there is no hyperfine structure. The effect of the hyperfine 
structure on the polarization is most striking in the calculations 
for the maser emission that represents the central (or systemic) 
features of NGC4258. Information about magnetic fields is inferred
from observations for polarized maser radiation and bears on the structure 
of accretion disks.
\end{abstract} 


\section{Introduction}

Magnetic fields are likely to play a key role in the structure 
of accretion disks. In recent years, the idea that the long-sought 
physical mechanism for the viscosity in accretion disks involves 
turbulent magnetic fields has received special attention (e.g., 
Balbus \& Hawley 1998). The sub-parsec accretion disk at the nucleus of the galaxy 
NGC4258 provides a nearly ideal example of an accretion disk 
that is surrounding a massive, compact object (Watson \& Wallin
1994; Miyoshi et al. 1995)---presumed to be a black hole. Because 
this disk is observed at radio wavelengths and in a spectral 
line that is masing (the $6_{16} - 5_{23}$ transition of the water 
molecule), uniquely 
refined data about the properties of this disk have been obtained 
by utilizing the high resolution capabilities of very long baseline 
interferometry. Polarization of the maser radiation can potentially 
provide information about the magnetic field and is observed 
in the radiation from water masers located in our own Galaxy. 
The observations of NGC4258 to date have yielded only upper limits 
for the linear (Deguchi, Nakai, \& Barvainis 1995; Herrnstein 
et al. 1998) and circular (Herrnstein et al. 1998) polarization. 
These observations have focussed on the emission from the side 
of the disk which occurs in narrow spectral lines that are relatively 
weak. In contrast, the maser flux from directions toward the 
center of the disk (the systemic emission) is much greater than 
that from the sides and would seem to offer the possibility for 
measurements that are more sensitive (e.g., Herrnstein et al. 
1998). Instruments which are under consideration for the future 
also have the potential for increased sensitivity to the polarization 
of the water masers (e.g., Moran 2000).

In previous calculations (Wallin, Watson, \& Wyld 1998, 1999; 
hereafter Papers I \& II), turbulence in the velocities that 
is plausible for an accretion disk was found to lead naturally 
to spectra that are similar to the observed spectra of NGC4258. 
Utilizing these spectra as a basis, we now investigate the relationship 
between the circular polarization of the maser radiation and 
magnetic fields that are likely to be present in the disk. Keplerian 
velocity gradients and turbulence in the velocities determine 
where the maser amplification mainly occurs. The magnetic field 
changes along the path of the radiation as a result of the turbulence. 
Simplifications that are valuable in interpreting the circular 
polarization for a two-level masing transition are no longer 
valid because the $6_{16}-5_{23}$ transition of the water molecule in 
astrophysical 
masers probably is a blend of three hyperfine components 
(Nedoluha \& Watson 1991). The circular polarization of the radiation 
from galactic $6_{16}-5_{23}$ water masers has been measured by Fiebig \& 
G\"usten (1989), and subsequently by others. However, instrumental effects 
prevent the line profile for the circular polarization from being 
determined well enough to distinguish the profile that is predicted 
for a blending of hyperfine components from that for a 
single two-level transition, which is antisymmetric about line 
center (R. M. Crutcher 2001, private communication). Observational 
support for blending does exist from the statistics of the spectral 
linebreadths for the flux of Galactic water masers (Gwinn 1994). In 
calculations for the maser inversion (e.g., Anderson \& Watson 1993),
a number of radiative and collisional transitions are found to influence
the populations of the $6_{16}$ and $5_{23}$ states. All such calculations to
date obtain population inversions for these and the several other transitions of water that have been observed to be masing without considering the hyperfine structure. The breadths of the infrared spectral lines involved in 
the pumping also are much greater than the hyperfine splitting. It 
is thus difficult to imagine that the maser pumping might select only 
one of the $6_{16}-5_{23}$ hyperfine transitions. In Papers I and II, we do simplify the calculations by ignoring maser saturation. In those investigations 
and elsewhere (Watson \& Wyld 2000), we have reasoned that the 
masers of NGC4258 are (at most) only moderately saturated because 
of their small linebreadths. A key idealization in the formulation 
of Papers I and II which is also utilized here is that the maser 
opacity is constant. The variations in the calculated spectra 
and spatial images are then entirely a result of changes in the 
velocity caused by the turbulence and by the Keplerian rotation. 
Finally, the only cause for circular polarization that will be 
considered here is the Zeeman effect---which can be treated in 
the limit in which the Zeeman splitting is much smaller than 
the spectral linebreadth. The presumed absence of significant 
maser saturation and the observed absence of significant fractional 
linear polarization reduce the likelihood for circular polarization 
due to other causes that have been investigated (Nedoluha \& 
Watson 1994; Wiebe \& Watson 1998).

For the three hyperfine components that we are considering, the
Zeeman splitting ($g\Omega$) in frequency units ranges from
approximately $10^4 B({\rm G})$~s$^{-1}$ for the $F=7-6$ transition
to approximately $10^3 B({\rm G})$~s$^{-1}$ for the $F=5-4$
transition. Consideration of the equipartition of energy and of
the magnitudes of the magnetic fields that have been detected in
galactic, 22 GHz water masers  indicate that  the magnetic fields
in the NGC4258 masers most likely are at least $10^{-2}$~G. A
decay rate $\Gamma\approx1$~s$^{-1}$ ordinarily is adopted for
the $6_{16}$ and $5_{23}$ states in astrophysical water masers.
Hence, simplifications in the calculation which are appropriate
when $g\Omega$ is much greater than $\Gamma$ and the rate for
stimulated emission (which is assumed to be less than $\Gamma$
since the masers are treated as unsaturated) are reasonable. In
this regime, there are no phase relations between the magnetic
substates. Only ordinary populations of the magnetic substates
then enter into the calculations. These lead to the constant maser
opacity for unsaturated masers that is adopted here. In contrast,
elements of the quantum mechanical density matrix are needed when
the rate for stimulated emission approaches and exceeds $g\Omega$.

The calculational methods are outlined in Section 2, results 
for the polarized spectra from the sides and from directions 
toward the center of the disk are presented separately in Section 
3, and further discussion of the significance of the results 
is given in Section 4.

\section{Basic Methods}

Except for the calculation of the circular polarization 
of the radiation, the basic methods are exactly the same as in 
our previous calculations where they are described in more detail 
(Papers I and II). Representative turbulent velocity fields (``statistical 
realizations'') are created in rectangular volumes by statistical 
sampling. To accomplish this, a statistical sample of the Fourier 
amplitudes of the velocity field is chosen from Gaussian distributions 
for the amplitudes and a Kolmogorov-like power spectrum. These 
methods are standard (e.g., Dubinski, Narayan, \& Phillips 1995). 
The rectangular volumes of turbulent velocities are imagined 
to be located at the sides or along the line to the center of 
the galaxy as viewed by a distant observer in the plane of the 
disk (see Figure 1). The bulk velocity of the masing gas at a 
location is then taken to be the sum of the turbulent velocity 
and the circular, Keplerian velocity appropriate for the disk 
of NGC4258 at that location. The masing is treated in the unsaturated 
limit as discussed in the Introduction. In other contexts, we 
have found that modest saturation does not alter significantly 
the relationship between the circular polarization and the magnetic 
field for water masers (Nedoluha \& Watson 1992). Incorporating 
maser saturation would make the calculation far more difficult. 
In any case, it is desirable to understand the relationship between 
the circular polarization of the maser radiation and the magnetic 
field in a turbulent, Keplerian disk in the simpler, unsaturated 
limit before contemplating the possible effects of maser saturation 
on this relationship. For the distant observer in the plane 
of the disk, the observed maser flux at Doppler velocity $v$ in 
terms of the maser optical depth $\tau(v,b,z)$ [defined here as a positive 
quantity] for a ray of radiation at impact parameter $b$ measured 
from the center of the disk and at distance $z$ measured from the 
midplane of the disk is then (e.g., Paper I)
\begin{equation}
S(v) = {2\nu^2kT_{\rm c}\over c^2D^2}
\int db\,dz \{\exp [\tau (v,b,z)] - 1\}
\equiv {2\nu^2kT_{\rm c}\over c^2D^2}HR_0{\cal S}(v).
\end{equation}
Here, $D$ is the distance from the observer to the galaxy, $T_{\rm c}$ is 
the temperature of the background continuum radiation, $H$ is the 
thickness of the disk and $R_0$ is the inner radius of the masing 
region. We adopt $R_0 = 4\times 10^{17}$ cm and $H = 4\times10^{15}$ cm for 
NGC4258 (e.g., Papers I \& II). 
The normalized flux ${\cal S}(v)$ is the factor by which the disk amplifies 
radiation at Doppler velocity $v$. In writing equation (1), the 
masing is assumed to be amplifying background continuum (as opposed 
to spontaneous) radiation. Since the simplifying idealization 
is being made (see Introduction) that the maser opacity $\kappa_0$ is 
constant, the optical depth for a single masing transition (a 
transition involving only two energy levels) can be expressed 
as 
\begin{equation} 
\tau(v,b,z) = \kappa_0 \int \exp(-v^2_1/v^2_{\rm th})\,ds
\end{equation}
where the integration is along the straight line path of the 
ray through and parallel to the plane of the disk. The Doppler 
velocities in equation (2) are related by 
\begin{equation} 
v_1 = v-v^{\rm K} - v^{\rm t}
\end{equation}
where $v^{\rm K}$ and $v^{\rm t}$ are the projections along the line of sight 
of the Keplerian and the turbulent velocities, respectively. 
To obtain the thermal velocity $v_{\rm th}$ for the water molecules, we 
adopt a temperature of 500~K as representative for the masing 
of water in astronomical environments. In effect, the foregoing 
simplifications reflect the idealization that the main cause 
for the spectral and spatial variations in the maser intensities 
are due to the changes in the velocity field. To designate the 
energy flux that is computed in the two energy level approximation 
using the optical depth from equation (2), we append the subscript 
``2'' to obtain $S_2(v)$ and ${\cal S}_2(v)$.

Representative turbulent magnetic fields are created in the rectangular 
volumes at the center and sides of the disk by exactly the same procedure 
as is utilized for the turbulent velocity fields. The 
random numbers that are used in the statistical sampling are, 
of course, not the same. The parameters that describe the turbulence 
in the magnetic fields---the Kolmogorov power spectrum and the 
largest wavelength in the power spectrum---are the same as for 
the velocity fields. The largest wavelength corresponds to the 
thickness $H$ of the disk. Since the circular polarization depends 
linearly on the strength of the magnetic fields, the Stokes-$V$ 
that is calculated by incorporating turbulent magnetic fields 
is thus proportional to the rms value $B_{\rm rms}$ of these turbulent 
magnetic fields. Numerical simulations of MHD turbulence in astrophysical 
disks do tend to indicate that the power spectra of the velocity and 
magnetic fields are similar (e.g., Brandenburg et al. 1995; Stone et al. 
1996), and thus that our procedure is a reasonable 
way in which to idealize turbulent magnetic fields in gaseous, 
Keplerian disks in astrophysics. The two power spectra also are 
similar in MHD simulations for non-rotating media that are intended 
to represent interstellar clouds (e.g., V\'azquez-Semadeni et al. 
2000).

When the Zeeman splitting is much smaller than the linebreadth 
of a spectral line and the maser is unsaturated as is the case here,
it is an excellent approximation to
assume that the local contribution to the Stokes-$V$ intensity for an 
individual ray is proportional to the derivative with 
respect to Doppler velocity of the source function for the intensity. 
Since the fractional circular polarization also is small as in 
this regime, for a two-level transition
\begin{equation} 
V_2(v,b,z) = -2\kappa_0I(v,b,z)p \int (v_1/v^2_{\rm th})
B_{\rm s}\exp(-v_1^2/v^2_{\rm th})\,ds
\end{equation}
where $B_{\rm s}$ is the component of the magnetic field that is parallel 
to the line of sight and $p$ is a constant that involves the product
of the magnetic moment and the transition probability. Also, 
the intensity of the individual ray is $I(v,b,z)=I_{\rm c}\exp[\tau (v,b,z)]$ 
where $I_{\rm c}$ is the intensity of the background continuum radiation. The 
integrand in equation (4) is a function 
of $b$ and $z$, as well. When the magnetic field is constant, equation (4) 
becomes
\begin{equation} 
V_2(v,b,z)=pB_{\rm s}\,{\partial I\over\partial v}(v,b,z).
\end{equation} 
Thus, investigators commonly seek to relate Stokes-$V$ to the derivative of 
the intensity when the Zeeman splitting is much smaller than the spectral 
linebreadth. The observed flux of circularly polarized radiation at a 
specific Doppler velocity is then the sum, analogous to that in equation 
(1), of the contributions of all rays to the Stokes-$V$ flux at this Doppler
velocity
\begin{equation} 
S_{V2}(v)=\int db\,dz V_2(v,b,z)\equiv {2\nu^2kT_{\rm c}\over c^2D^2}
HR_0{\cal S}_{V2}(v)
\end{equation}
in which a normalized Stokes-$V$ flux ${\cal S}_{V2}(v)$ is defined in analogy 
with the normalized energy flux in equation (1). We have emphasized in the 
Introduction that there is no basis for treating the $6_{16}-5_{23}$ masing 
spectral line as a two level transition and that the three strongest 
hyperfine components should always be treated together. For the three 
hyperfine components together, the foregoing equations can be expressed as 
(see, e.g., Nedoluha \& Watson 1992)
\begin{eqnarray}
\tau(v,b,z)=\kappa_0 \int \{0.385\exp[-(v_1 - 0.45\,{\rm km\,s}^{-1})^2
/v^2_{\rm th}]+0.324\exp[-v^2_1/v^2_{\rm th}]+\nonumber\\ 
+ 0.273\exp[-(v_1 + 0.58\,{\rm km\,s}^{-1})^2
/v^2_{\rm th}]\}\,ds 
\end{eqnarray}
and 
\begin{eqnarray}
V(v,b,z)= - 2\kappa_0I(v,b,z)p
\int 0.385B_{\rm s}\left\{[(v_1-0.45\,{\rm km\,s}^{-1})/v^2_{\rm th}]
\exp[-(v_1-0.45\,
{\rm km\,s}^{-1})^2/v^2_{\rm th}]\right.+\nonumber\\
+ 0.52(v_1/v^2_{\rm th})\exp(-v^2_1/v_{\rm th}^2) 
\left.+ 0.052[(v_1 + 0.58\,{\rm km\,s}^{-1})/v^2_{\rm th}]\exp[-(v_1 + 
0.58\,{\rm km\,s}^{-1})^2/v^2_{\rm th}]\right\}ds.
\end{eqnarray}
To compute the flux $S(v)$ when all three hyperfine transitions 
are considered, $\tau$ from equation (7) is inserted into equation 
(1). Likewise, $S_V, V(v,b,z)$ of equation (8) and the normalized 
Stokes-$V$ flux ${\cal S}_V(v)$ are related by an equation exactly 
analogous to equation (6).

Integration over $z$ and $b$ in the foregoing is appropriate 
for comparisons with data for NGC4258 because the spatial resolution 
of the observations is not sufficient to resolve the thickness 
of the disk nor the range of impact parameters from which radiation 
emerges at a specific Doppler velocity.

\section{Results of Calculations}

\subsection{Maser Radiation from the Sides of the Disk}

The appearance of the spectrum for the emission from the sides 
of the disk varies somewhat from one statistical realization 
of the turbulent velocity field to another (Paper II). Statistical 
realizations are different because of the initial (or seed) number 
given to the random number generator. As a representative basis 
for examining the polarization, we utilize a realization for 
the velocity fields that leads to a spectrum (panel a of Figure 
2 in Paper II) for $S(v)$ that is the most similar to the observed 
spectrum of NGC4258---that is, a spectrum that consists of several, 
mostly well separated, narrow lines (1 to 2 km s$^{-1}$ FWHM) with 
peak fluxes that are comparable and are dispersed in Doppler 
velocities over some 200 to 300 km s$^{-1}$. This spectrum is shown 
in Figure 2 in terms of the normalized flux density ${\cal S}(v)$ obtained 
by including all three hyperfine components in the calculation for the 
optical depth $\tau$. As discussed in Paper II, the opacity parameter is 
chosen to yield peak fluxes 
for the spectral lines that are similar to the fluxes that are 
observed. We have reasoned in Paper II that $T_{\rm c}$ for the masers at the sides of the disk is likely to be 10 to 100 degrees K. We also reason in Paper II that the magnitude of the 
opacity that we utilize is plausible for the $6_{16}-5_{23}$ water masers. There are only modest differences in this spectrum when the optical depth $\tau$ is 
computed from equation (2) for a single masing transition instead of with 
all three hyperfine components. That is, ${\cal S}_2(v)$ and ${\cal S}(v)$
are similar in character. Comparisons 
are shown in the upper six panels in Figure 3 for the individual 
features labelled (a)--(f) in Figure 2. In the middle panels in 
Figure 3, we compare the derivative of the flux
$\partial {\cal S}/\partial v$ for these spectral features with the 
Stokes-$V$ flux ${\cal S}_V(v)$ due to a constant magnetic field when 
the three hyperfine components are included in the calculations 
for both as in equations (7) and (8). The purpose of this comparison 
is to demonstrate that, {\em even for a magnetic field that can be 
treated as constant throughout the masing region}, the actual 
spectrum for Stokes-$V$ should be quite different from the derivative 
of the energy flux because of blending of the hyperfine components. 
The hyperfine nature of the $6_{16}-5_{23}$ transition often is ignored, 
and the Stokes-$V$ flux is assumed to be proportional to the product 
of the strength of the magnetic field $B_{\rm s}$ and the ``derivative 
spectrum'' $\partial {\cal S}/\partial v$ as is valid for a two-level transition. 
A value or upper limit for $B_{\rm s}$ could then be inferred. The comparisons 
in Figure 3 emphasize, however, that this procedure is not valid---as 
demonstrated previously (Nedoluha \& Watson 1992) for a medium 
without bulk motions. Note that, in contrast to the derivative 
spectrum, the Stokes-$V$ spectrum does not tend to change sign 
as one moves across the spectral line. In the bottom panel in 
Figure 3, we show the normalized Stokes-$V$ flux ${\cal S}_V(v)$
that is obtained for the features in
the top panels when the magnetic fields are not constant, but 
are entirely turbulent and are specified as we have discussed 
in Section 2. To obtain an indication of the possibilities, we 
show the results of two separate computations for two independent 
statistical realizations of the magnetic field. In some cases 
(see features a, c, and f) variations occur that are suggestive 
of the derivative spectrum and can thus be misleading when interpreting 
the observations if they are taken as evidence that the $6_{16}-5_{23}$ 
masing involves only a single transition and not all three hyperfine 
components. 

To provide a further indication of how spectra of the circular polarization
can appear when the magnetic field is turbulent, the Stokes-$V$ flux
${\cal S}_{V2}(v)$ is presented in Figure 4 for two of the features in
Figure 2 {\em when the $6_{16}-5_{23}$ masing is treated as a single,
two-level transition}. It is compared with the derivative spectrum
$\partial {\cal S}_2/ \partial v$ for these features. These spectra are
representative. The Stokes-$V$ and the derivative spectrum can be quite
similar as in panel (b), though they are more likely to be somewhat
different in appearance. Two-level spectra for Stokes-$V$ which appear
similar to the Stokes-$V$ that are computed with the three hyperfine
components also occur (panel a). No change in sign occurs for these spectra
as one moves across the line profile.

Representative examples are shown in Figure 5 for the variation of
the turbulent magnetic field component $B_{\rm s}$ along the path of a ray and
for variation of the maser opacity $R_{\rm 0}\partial\tau/\partial s$. The
Doppler velocities for these two rays are at the peaks of two of
the spectral features in Figure 2. For one ray, the contributions
to the optical depth are spread relatively uniformly along its path,
whereas the contributions can be viewed as arising from two ``clumps''
for the other ray. The latter configuration tends to support the idea
that ``aligned masers'' (Deguchi \& Watson 1989) play an important role
for the emission near the planes of Keplerian and other disks.

The circular polarization of spectral lines at radio frequencies often
is quite weak so that it is difficult to delineate a line shape
accurately enough to infer $B_{\rm s}$ by fitting the observed Stokes-$V$
to the derivative spectrum---even assuming a two-level transition
and that $B_{\rm s}$ is a constant throughout the region in which the spectral
line is created. Under such circumstances and when the Zeeman splitting
is much smaller than the spectral linebreadth, a useful relationship
from which an excellent approximate value $B_{\rm i}$ can be inferred
for the magnitude of $B_{\rm s}$ is
\begin{equation}
B_{\rm i}=(S_{V,\max}-S_{V,\min})\Delta v/2AS_{\max}
\end{equation}
where $S_{V,\max}$ and $S_{V,\min}$ are the maximum and minimum values
of Stokes-$V$ across the spectral line, $\Delta v$ is the Doppler breadth
(FWHM) of the spectral line, $S_{\max}$ is the peak intensity in the
spectral line, and $A$ is a constant that involves the magnetic moment
and the transition probability of the specific molecular transition.
Equation (9) was originally recognized to be useful for interpreting
non-masing spectral lines. Somewhat surprisingly, equation (9) also has
been found to be a good approximation for spectral lines that are
masing strongly, despite the ``exponential'' nature of maser amplification
(Fiebig \& G\"usten 1989). Even when the three hyperfine components of the
$6_{16}-5_{23}$ transition of water are blended and the maser is moderately
saturated, equation (9) is still useful (Nedoluha \& Watson 1992). In the
calculations for Figure 3 in which the magnetic field is constant and the
hyperfine components are blended, we also find that equation (9) is an
excellent approximation. As found by Nedoluha \& Watson (1992), the best
value for the inferred magnetic field when the blending of the hyperfine
components is included is about two-thirds of that given by equation (9).
To understand the statistical relationship between the observed circular
polarization and the turbulent magnetic field that actually exists in
the masing disk, we also will utilize equation (9) as a benchmark.

Since the magnetic field in our calculations is entirely turbulent,
the circular polarization of a ray of radiation in a spectral line
ideally should disappear as the size of the masing region becomes very
large in comparison with the scale length for variations of the turbulent
magnetic field, assuming a constant opacity parameter and no velocity
gradients. Because the maser amplification occurs within a finite distance
as a result of the Keplerian velocity gradients, the circular polarization
due to turbulent magnetic fields that we have computed for Figure 3 and
for other figures is not zero. Since this circular polarization depends
upon the statistical variations of the magnetic field, it can be expected
to vary in a statistical manner from one ray to another. To begin to see
how the observations can be related to the turbulent magnetic field, we
examine the relationship between the $B_{\rm i}$ that are determined from equation
(9) and the actual strength of the turbulent magnetic field $B_{\rm rms}$. For
this, the spectral features in Figure 2 are again utilized. A histogram
of the ratio $B_{\rm i}/B_{\rm rms}$ obtained from the spectra in the bottom panels of Figure 3 is shown in
Figure 6. In addition to those shown in Figure 3, Stokes-$V$ spectra are
computed for other realizations of the magnetic field to enlarge the sample
size for creating the histogram. As defined in equation (9), $B_{\rm i}$
is always positive. The statistical distribution that is indicated for actual
magnetic fields is, of course, the same for both positive and negative values.
For inferring $B_{\rm i}/B_{\rm rms}$ the factor $A$ in equation
(9) is taken to be that which is appropriate when the masing is due to the
strongest hyperfine transition ($F=7$ to 6) alone. Specifically, $A=2.87p$.
To further assess the usefulness of this ratio when the magnetic fields are
turbulent, we examine its sensitivity in Figure 7 to the peak flux in the
spectral feature in which the ratio is being determined. The evidence
in Figure 7 indicates that in the presence of turbulent velocity and
magnetic fields, the ratio still is insensitive to the strength of the
spectral line. The significance of Figure 6 is the following. Suppose
that observational data for Stokes-$V$ are obtained for a number of features
at the sides of NGC4258 and are interpreted on the basis of equation (9)
assuming a two-level ($F=7$ to 6) transition for the masing. The histogram
in Figure 6 indicates that the median value of the magnitudes of the
magnetic fields that are inferred will be about one-third of $B_{\rm rms}$
for magnetic fields in the disk that are entirely turbulent. If only upper
limits can be obtained, the upper limit for $B_{\rm rms}$ will then be about
three times median of the upper limits for the observed magnetic fields
that are obtained from equation (9). Clearly this histogram depends upon
the choice for the largest scale lengths in the distribution for the power
spectrum of the Fourier components. We have reasoned that the choice made
here provides a representative example. In view of the limited
observational efforts to date and ambiguity about the exact description
of the medium, a more extensive investigation of how this histogram
depends upon the specific nature of the turbulent medium of a disk
does not seem to be warranted at this time.

\subsection{Maser Radiation from Directions toward the Center of the Disk}

The general appearance of the spectrum of the maser radiation from
directions toward the center of the disk (the systemic emission) is less
sensitive in our calculations to the specific statistical realization of
the velocity field than is the spectrum computed in the foregoing Section
for the emission at the sides of the disk. It is noteworthy that the
appearance of the spectrum of the central emission from NGC4258 has varied
considerably over the decade or so in which it has been observed. The
variability of this central emission in comparison with that from
the sides of the disk is not surprising (e.g., Paper II). Comparison
of the top panels in Figures 8 and 9 indicate that, as for the spectrum
for the sides of the disk, the appearance of the spectrum for the energy flux
is relatively insensitive to whether all three hyperfines are included
in the calculation. The normalized fluxes in these figures are much smaller than those for the masers at the sides of the disk because the central masers are believed to be amplifying a bright continuum source with $T_{\rm c}$ of about
$10^{6}$ to $10^{7}$ degrees K. We have reasoned in Papers I and II that magnitudes of the normalized fluxes in Figures 8 and 9 are then appropriate for comparisons with the observations. When the normalized spectrum for the circular
polarization resulting from the turbulent magnetic field is compared
with the derivative spectrum in the second and third panels of Figure 8
for calculations that properly include the blending of hyperfine components,
it is evident that they are dissimilar. Clearly, an analysis in which the
observed Stokes-$V$ spectrum is fit to the derivative spectrum to obtain
the line-of-sight component of the magnetic field is inappropriate. This is
expected (as it was for the side features) since we already know that
such comparisons are invalid when all three hyperfine components are
properly included, even when there is no turbulence in the magnetic field.
What may be surprising is that in Figure 9 where the hyperfine structure is
ignored, there also is little if any similarity between the derivative and
Stokes-$V$ spectra. That is, changes in the magnetic field along the
line of sight are sufficient to alter completely the appearance of
Stokes-$V$ even though the source function for Stokes-$V$ at any location
along the path of the ray is proportional to the derivative of the source
function for the energy flux at that location. To extract information from
our calculations which may be useful to obtain a measure of the strength
of the irregular magnetic field from potential observations, we have
computed the standard deviation of the ratio
$(v_{\rm th}/pB_{\rm rms}){\cal S}_V(v)/{\cal S}(v)$ over the range
of Doppler velocities in the central emission. That is, we have computed
the standard deviation for the spectrum  in Figure 8 of the ratio of the
flux in panel (c) to that in panel (a). From equation (8), we suspect that
this quantity may be a measure of $B_{\rm rms}$ that is insensitive
to uncertainties in the calculation other than the correlation length
of the turbulence. For the realization on which Figure 8 is based, as well
as for a second representative realization, this standard deviation
is approximately 0.5. The analogous calculation for the same statistical
realizations, but for a two-level masing transition, yields standard
deviations that are approximately one. The mean value of the ratio should
be close to zero. For both types of transitions, we compute means that
are much smaller than the standard deviations. To obtain an indication
of the sensitivity of this standard deviation to uncertainties in the
calculation, we have performed computations with a larger value for
the maser opacity so that the normalized flux is increased by a factor
of ten. The standard deviations are increased by approximately
fifty percent.

The bottom panels in Figures 8 and 9 show Stokes-$V$ spectra for a magnetic
field that is constant and entirely in the azimuthal direction---that is,
in circles about the center of the disk. Because of the rotation of the
disk, an averaged magnetic field with this geometry is expected. No other
magnetic field---including a turbulent magnetic field---is present in this
case. Whereas in Figure 9, the spectrum for the two-level transition
exhibits sign reversals on scales of 1--2 km s$^{-1}$ similar to those of
the derivative spectrum as is expected, the spectrum in Figure 8 with
blended hyperfine components is quite different. Stokes-$V$ at negative
Doppler velocities are entirely of one sign, and at positive velocities
Stokes-$V$ is entirely of the opposite sign. The reversal in sign is,
of course, caused by the reversal in sign of the line-of-sight component
of the azimuthal magnetic field from one side of the disk to the other.
The absence of sign reversals on scales of 1--2 km s$^{-1}$ in this
spectrum is to be expected from the spectra in Figure 3 and in
earlier calculations (Nedoluha \& Watson 1992). We also show the Stokes-$V$
spectrum for a constant magnetic field along the line-of-sight in the
bottom panel of Figure 8. Since the angular portion of the disk that
contributes to the central emission of NGC4258 is quite small, this
spectrum is a good indication of Stokes-$V$ due to a constant, radial
magnetic field.

The ratio of Stokes-$V$ for constant $B_{\rm s}$ in Figure 8 (bottom panel)
to the energy flux for the same spectrum (top panel of Figure 8),
expressed as a dimensionless quantity, is given explicitly in
Figure 10. The rapid variations of this ratio over a few km s$^{-1}$ seem
to be due mainly to the shift between the peaks of ${\cal S}$ and
${\cal S}_V$ for the spectral features (e.g., Figure 3). When these
rapid variations are averaged, the ratio in Figure 10 changes only by
about fifty percent over the approximately 80 km s$^{-1}$ range of
the Doppler velocities within which the central emission from NGC4258
is observed. If sharp spectral features can be distinguished in the
central emission, equation (9) can be utilized to infer information
about the strength of the magnetic field.

\section{Discussion}

As a first requirement for assessing the circular polarization
based on calculations, it seems that the calculated spectrum for
the energy flux should resemble the observed spectra and that this
should be achieved within a plausible (albeit idealized) description
for an accretion disk and for the masing process. Turbulence that is
generally similar to what is being utilized here is believed likely
to be a key feature of accretion disks and our calculated spectra do
resemble those from the observations of NGC4258. The possibility
cannot be excluded, however, that there are alternative descriptions
which also can lead to spectra that are similar to what is observed.
In detail, the thickness of the disk (potentially including a significant
non-masing component) is poorly known (Papers I \& II; Desch, Wallin
\& Watson 1998). This thickness tends to determine the longest
wavelengths to which the turbulence extends, and is thus a key
consideration for changes in direction of the magnetic field along the
path of a ray of maser radiation. At a quantitative level, the results
of our calculations should thus clearly be taken only as indications.

The strongest conclusions from these calculations might have been
anticipated from our earlier calculations (Nedoluha \& Watson 1992)
for the circular polarization of water masers in the absence of
turbulence and Keplerian rotation. That is, because the
$6_{16}-5_{23}$ transition of water actually consists of three
hyperfine components, the spectral profile for Stokes-$V$ is not
proportional to the derivative (with respect to Doppler velocity)
of the energy flux even when the magnetic field is constant. When
the magnetic field varies along the path of a ray as it does
for turbulent fields, the resulting spectral profile for Stokes-$V$
may (as a statistical fluctuation) in some cases appear to have a
shape that is similar to that of the derivative of the energy flux.
Conversely, when the magnetic field is turbulent, the profile for
Stokes-$V$ for a two-level transition (which has no hyperfine structure)
can be completely different from the derivative of the observed energy
flux. Because of the sign reversals of a turbulent magnetic field
that occur along the path of the radiation, the resulting circular
polarization of the $6_{16}-5_{23}$ masing transition ordinarily will
be reduced by an amount that will vary from one spectral line to
another. For the turbulent magnetic fields that we adopt as
representative, the magnetic field strengths that would be inferred
from the radiation emitted at the sides of the disk by utilizing
standard methods [equation (9)] will typically be about one-third
of the rms value of the turbulent magnetic field (Figure 6). When
the magnetic field does not change sign along the path of the radiation,
the resulting Stokes-$V$ for the $6_{16}-5_{23}$ transition will be mostly
of one sign across the line profile. As a consequence, Stokes-$V$ for
the central (or systemic) maser emission that is due to an azimuthal
magnetic field will only change sign at the center of the
approximately 80 km s$^{-1}$ range of Doppler velocities over which
it is observed. For a radial magnetic field, Stokes-$V$ for the central
emission will be essentially of one sign. Greater sensitivity to Stokes-$V$
and hence to the magnetic field can be achieved in observations of
the central maser emission since the flux is much greater than that
in the features from the sides of the disk of NGC4258. However, the
azimuthal component of the magnetic field is likely to be the
strongest in a Keplerian disk. Whereas this component is essentially
parallel to the path of the radiation in the side features, its
projection along the line-of-sight is only a few percent of the
total azimuthal field for emission at Doppler velocities of a few
tens of km s$^{-1}$ relative to the systemic velocity of the galaxy.

Circular polarization of maser radiation where the Zeeman splitting is
much less than the spectral line breadth can be caused by mechanisms
other than the Zeeman effect (Nedoluha \& Watson 1994; Wiebe \& Watson
1998). Such mechanisms probably dominate for the circular polarization
of the circumstellar SiO masers. However, they do require that the
fractional linear polarization typically be greater than the
fractional circular polarization by a factor of ten or more. Though
perhaps not sensitive enough to exclude completely such possibilities, the
absence of detectable linear polarization at the level of a few percent
of the total flux does tend to support the premise that the
circular polarization which may be detected can be interpreted in terms
of the Zeeman effect to infer information about the strength of the
magnetic field in the disk. It is noteworthy that the spectral line
profile for Stokes-$V$ due to these non-Zeeman effects would probably
be similar to that of a two-level transition. In any case, the
observations can provide upper limits to the magnetic fields.

\acknowledgments

We are grateful to B. K. Wallin for helpful information about computing the spectra of the disk, and to H.W. Wyld for helpful discussions. This research has been supported in part by NSF Grant AST99-88104.

\clearpage

\begin{figure}
\epsscale{0.8}
\plotone{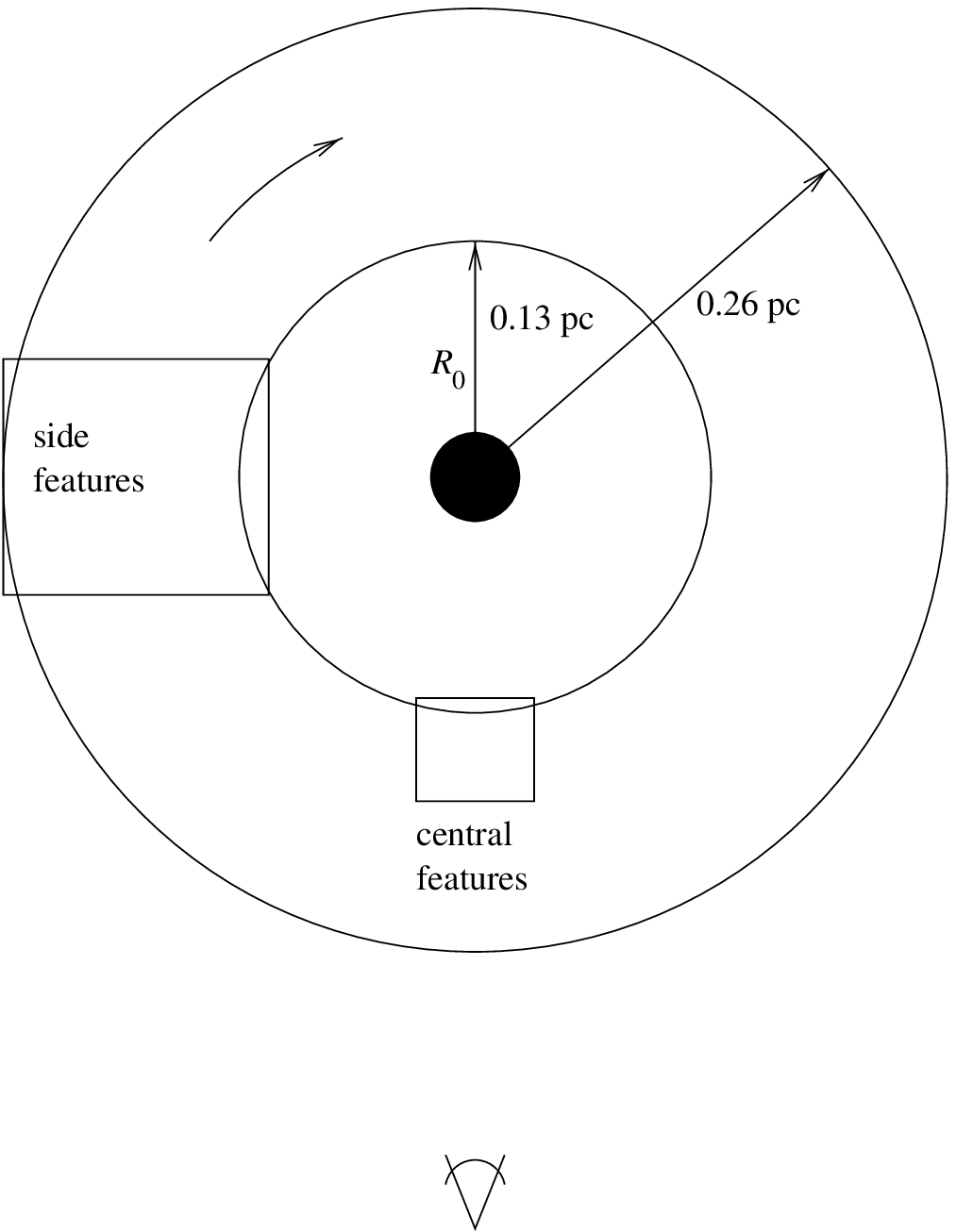}
\caption{%
Schematic diagram for a top view of the masing disk at the nucleus of the
galaxy NGC4258. The rectangles indicate the locations of the volumes
within which the turbulent velocities and magnetic fields are created
for computing the maser emission from the side and from the direction
toward the center of the disk. The observer is in the plane of the
disk. The rectangular volume shown at the center is not to scale and
extends only from $R_0$ to $1.05\,R_0$.\label{fig1}}
\end{figure}

\clearpage

\begin{figure}
\epsscale{1.}
\plotone{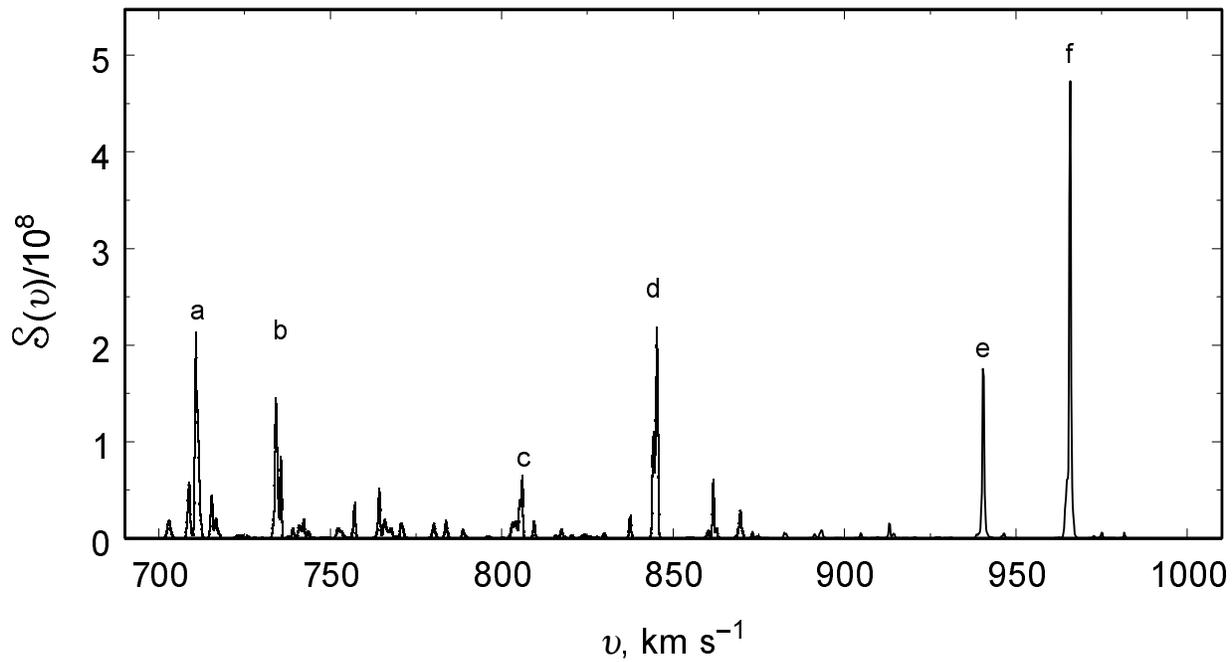}
\caption{%
A computed spectrum of maser radiation from the side of an
idealized Keplerian disk viewed edge-on. This spectrum is intended
to be representative of the emission of water masers from the sides
of the disk in NGC4258. The normalized flux ${\cal S}(v)$ is shown
as a function of the Doppler velocity $v$ (measured relative to
the 480 km s$^{-1}$ systemic velocity of NGC4258). All three
dominant hyperfine components of the $6_{16}-5_{23}$ transition
are included in the computations.\label{fig2}}
\end{figure}

\clearpage

\begin{figure}
\epsscale{0.8}
\plotone{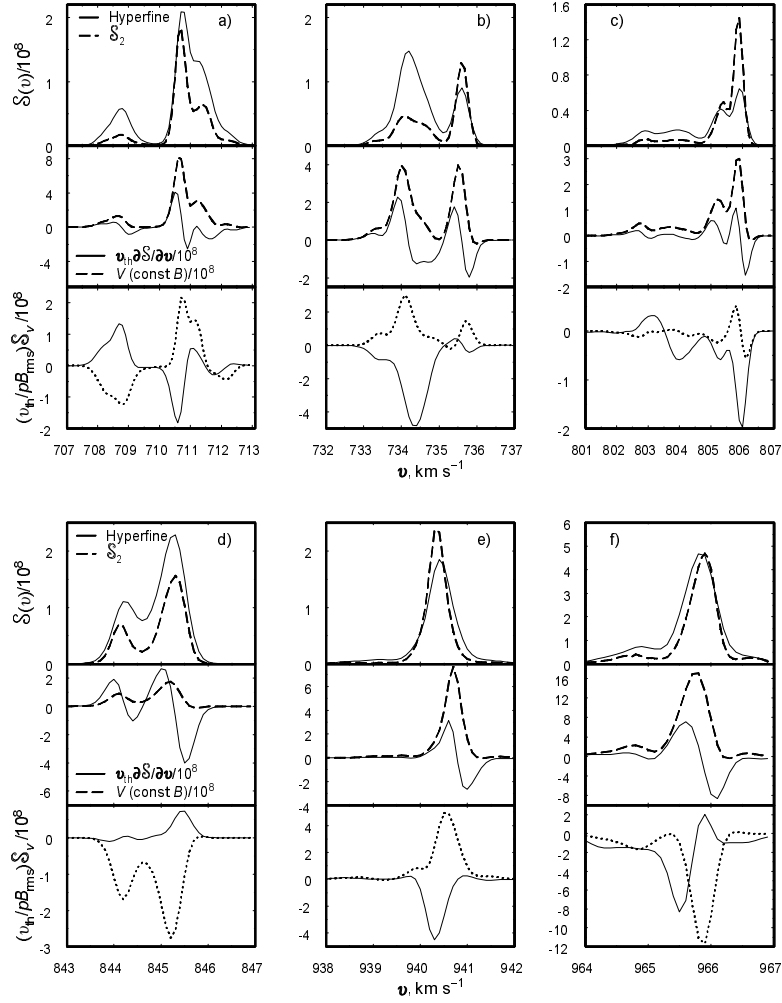}
\caption{%
Detailed spectra of individual features designated by the labels (a)--(f)
here and in Figure 2. The following quantities are shown for each
feature: (top panel) the normalized flux ${\cal S}(v)$ when all three
hyperfine transitions are included and the flux ${\cal S}_2$ when the transition
is assumed to occur between only two levels; (middle panel) the derivative
$v_{\rm th}\partial{\cal S}/\partial v$ and the normalized Stokes-$V$ flux
$(v_{\rm th}/pB_0){\cal S}_V(v)$ for a constant magnetic field $B_0$
along the line of sight---labelled as $V$(const $B$); and (bottom panel)
the normalized Stokes-$V$ flux $(v_{\rm th}/pB_{\rm rms}){\cal S}_V(v)$ for rms
turbulent magnetic fields $B_{\rm rms}$ along the line of sight. In the
bottom panel, Stokes-$V$ is shown for two different statistical
realizations of the turbulent magnetic field. Except for the dashed curve
in the upper panel labelled as ${\cal S}_2$, all the computations in
this Figure include the three dominant hyperfine components of the
$6_{16}-5_{23}$ transition of the water molecule.\label{fig3}}
\end{figure}

\clearpage

\begin{figure}
\epsscale{1.}
\plotone{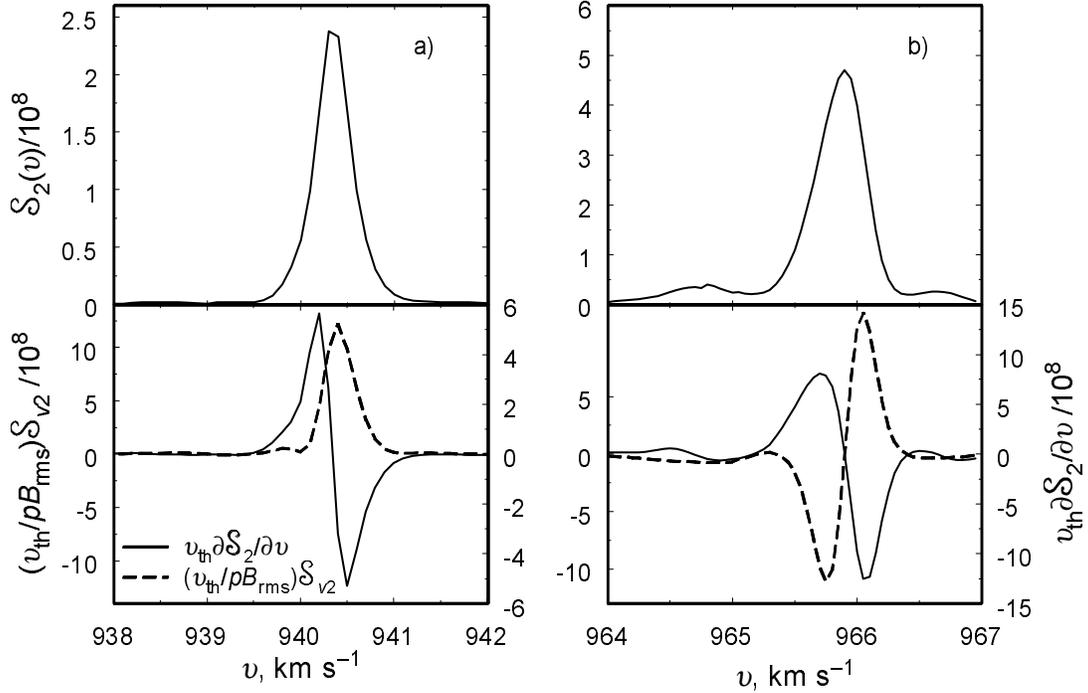}
\caption{%
Detailed spectra for the normalized flux ${\cal S}_2(v)$ {\em (top panels)}
for two representative features in Figure 2 obtained by treating the
transition as occurring between only two energy level. In the {\em bottom
panels}, the derivatives $v_{\rm th}\partial{\cal S}_2(v)/\partial v$ of the
spectra in the upper panels are compared  with examples of the normalized
Stokes-$V$ fluxes that are obtained for these features with rms turbulent
magnetic fields $B_{\rm rms}$.\label{fig4}}
\end{figure}

\clearpage

\begin{figure}
\epsscale{1.}
\plotone{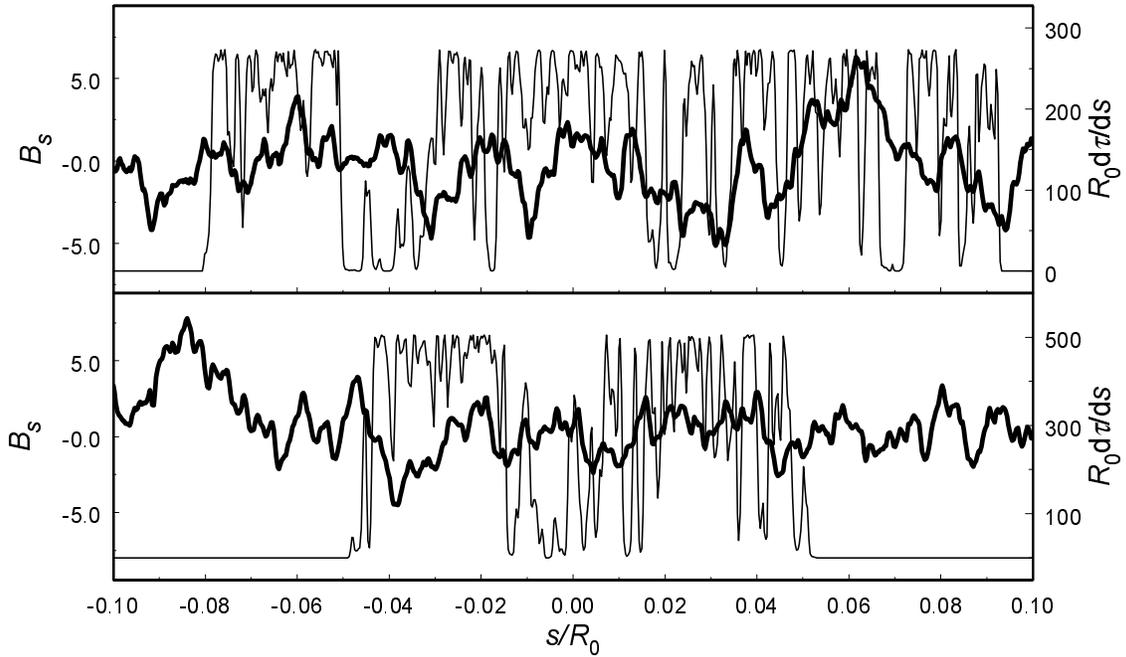}
\caption{%
The component of the turbulent magnetic field that is parallel to the line
of sight $B_{\rm s}$ (thick line) and the differential maser optical depth
$R_{\rm 0}\,d\tau/ds$ (thin line) as a function of distance along the paths of two
representative rays. These two rays are selected to have Doppler velocities
that are at the centers of two of the features in Figure 2.\label{fig5}}
\end{figure}

\clearpage

\begin{figure}
\epsscale{1.}
\plotone{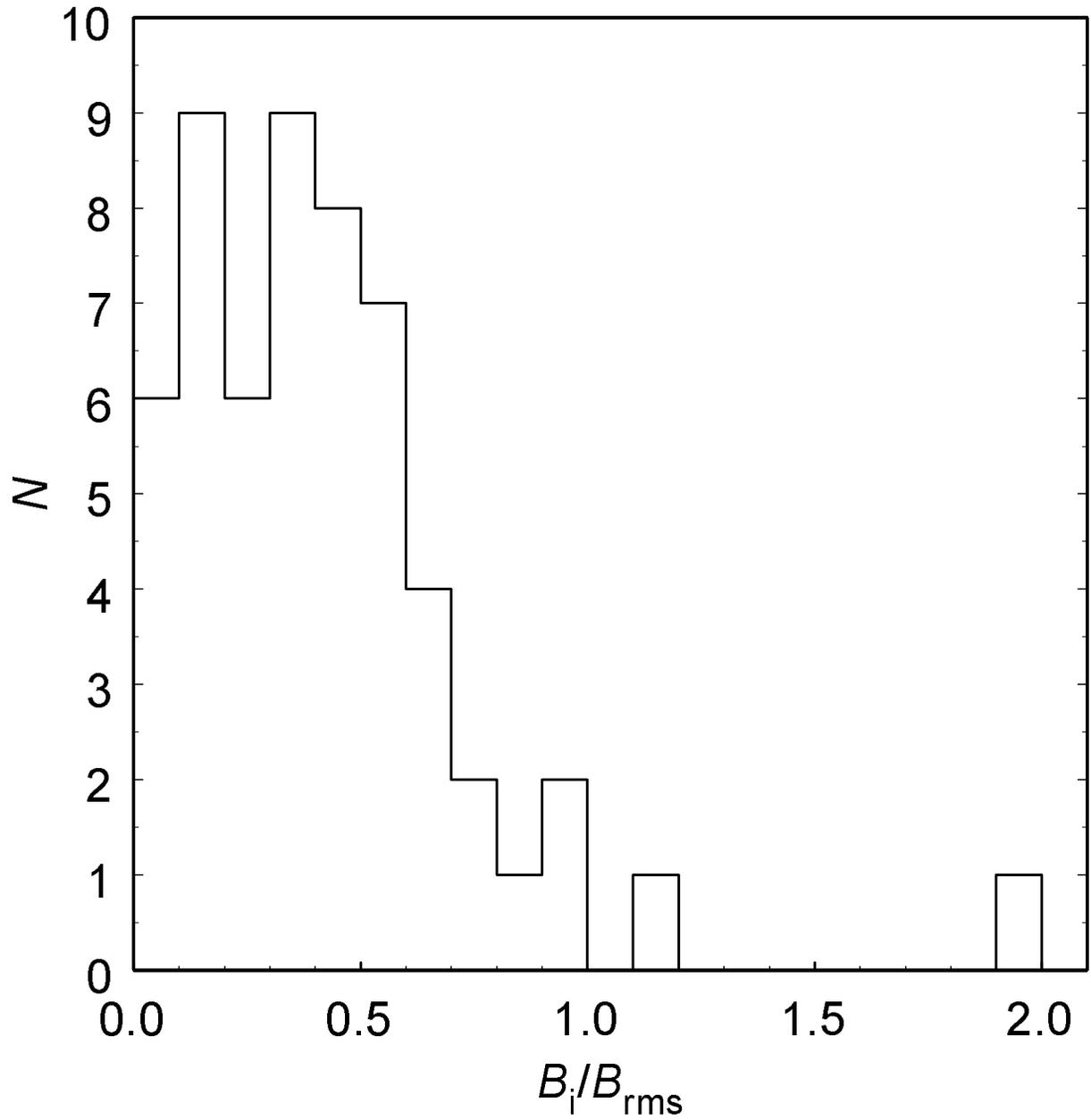}
\caption{%
Histogram of the ratio $B_{\rm i}/B_{\rm rms}$ obtained for the spectral
features in Figure 2 and Stokes-$V$ fluxes computed with several
statistical realizations for the turbulent magnetic field. All
the three dominant hyperfine components are included.\label{fig6}}
\end{figure}

\clearpage

\begin{figure}
\epsscale{1.}
\plotone{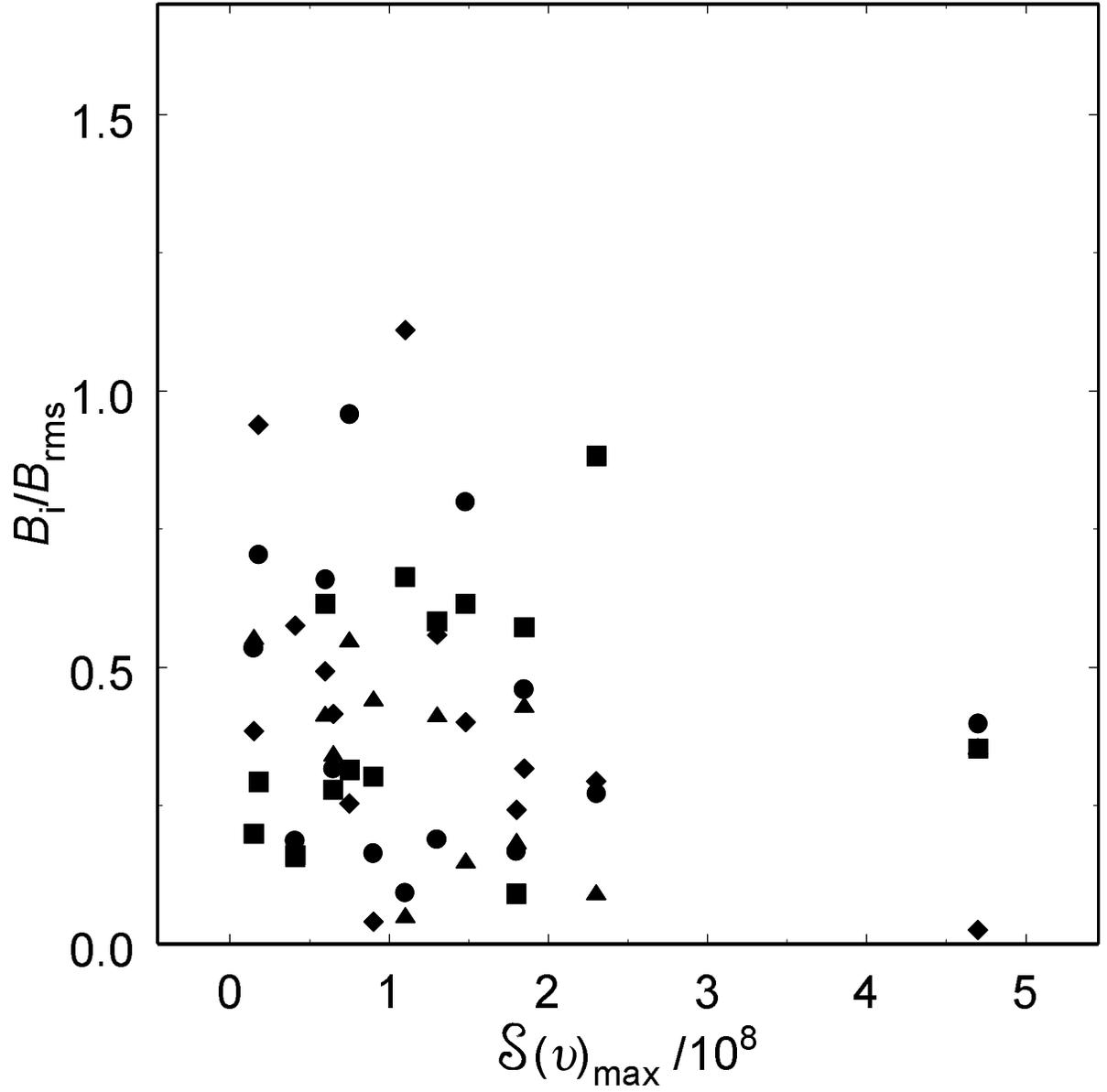}
\caption{%
The values for the ratio $B_{\rm i}/B_{\rm rms}$ utilized for the histogram
in Figure 6 as a function of the peak values ${\cal S}(v)_{\max}$
of the normalized flux in each ratio.\label{fig7}}
\end{figure}

\clearpage

\begin{figure}
\epsscale{0.8}
\plotone{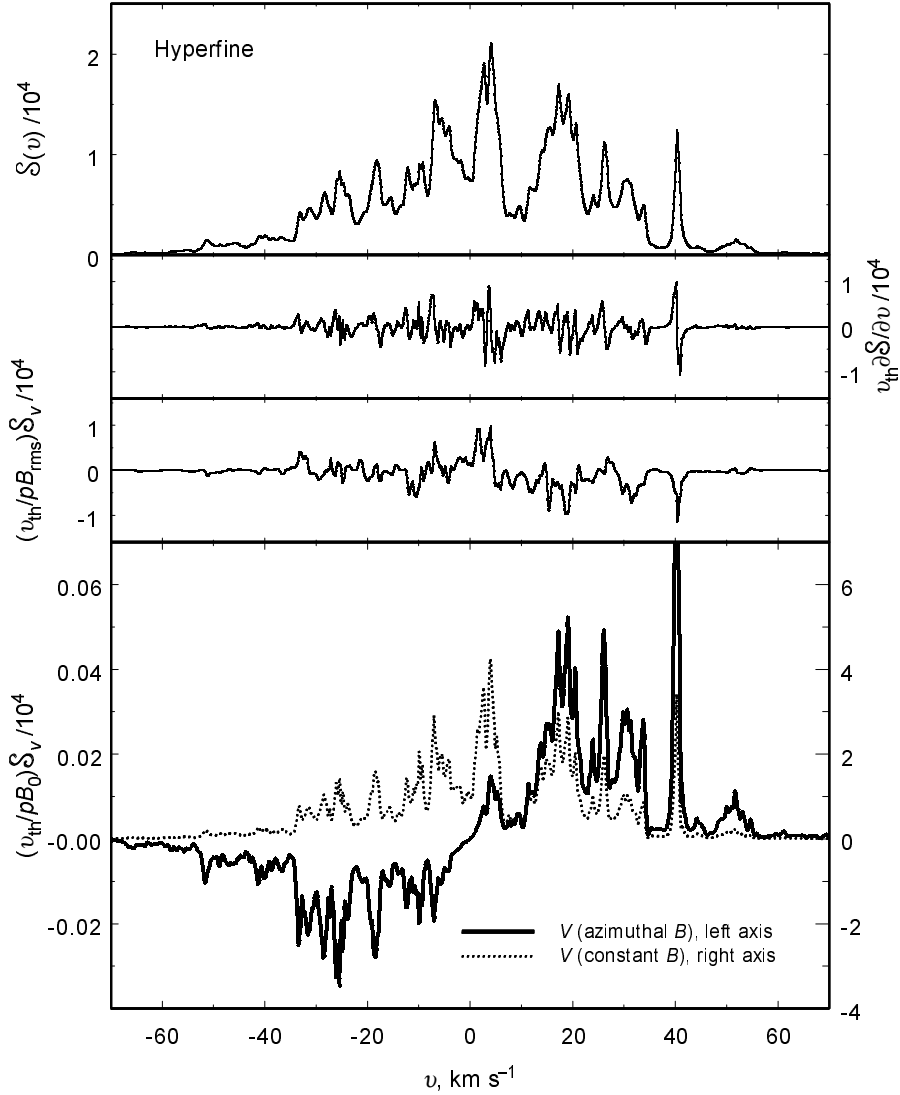}
\caption{%
Computed spectra for maser radiation from {\em central region} of an
idealized Keplerian disk viewed edge-on. These spectra are
intended to be representative of the emission near the systemic
velocity by water masers located along the line of sight to the
{\em center} of the disk in NGC4258. The spectra are shown as a function
of the Doppler velocity $v$ (measured relative to the 480 km
s$^{-1}$ systemic velocity of NGC4258). All three dominant hyperfine
components are included. {\em (top panel)} The normalized maser flux
${\cal S}(v)$. {\em (second panel)} The derivative
$v_{\rm th}\partial{\cal S}(v)/\partial v$ of the flux in the top panel.
{\em (third panel)} Normalized Stokes-$V$ flux $(v_{\rm th}/pB_{\rm rms})
{\cal S}_{V}(v)$ for the spectrum in the top panel due to a turbulent
magnetic field. The results of computations for two cases
(statistical realizations) of the turbulent magnetic field are shown.
{\em (bottom panel)} Normalized Stokes-$V$ flux $(v_{\rm th}/pB_0)
{\cal S}_{V}(v)$ due to a constant magnetic field $B_0$ (broken line),
and due to a magnetic field that is entirely in the azimuthal
direction about the center of the galaxy (solid line), for the
spectrum in the top panel.\label{fig8}}
\end{figure}

\clearpage

\begin{figure}
\epsscale{0.8}
\plotone{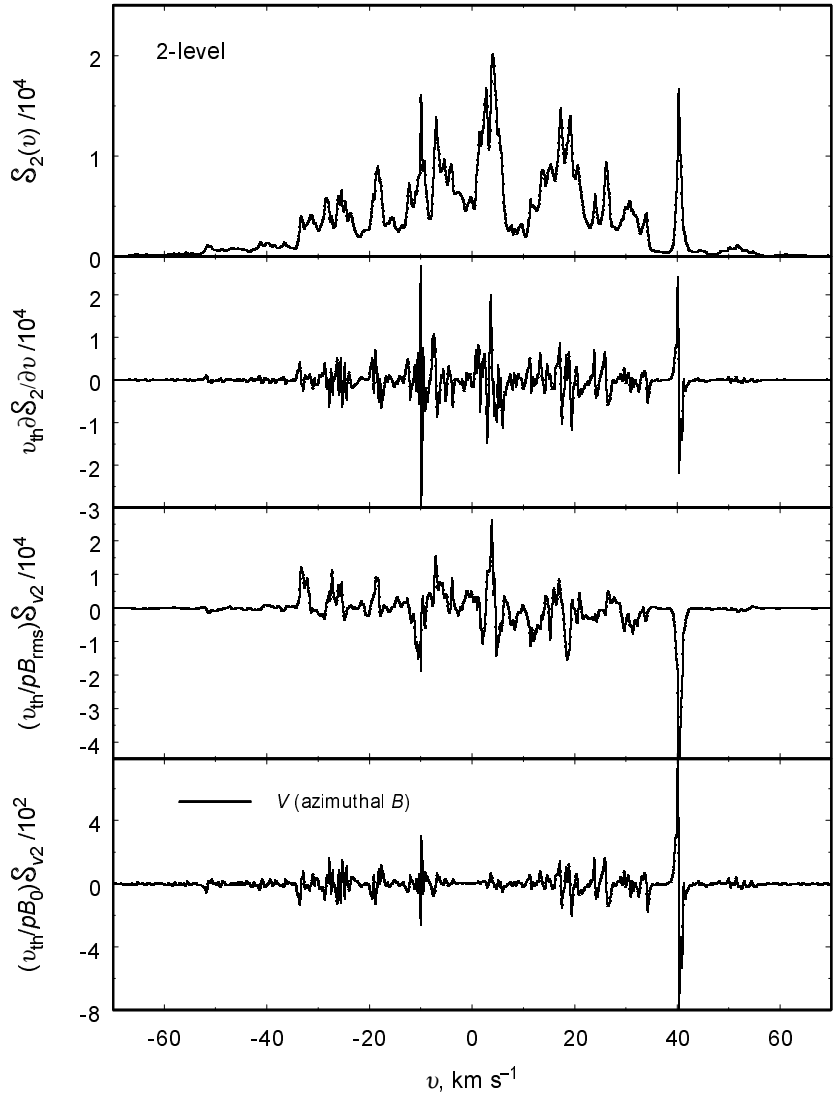}
\caption{%
Same as Figure 8, except that the masing is treated as due to only
a single, two level transition---instead of to the three hyperfine
transitions included in the computations for Figure 8.\label{fig9}}
\end{figure}

\clearpage

\begin{figure}
\epsscale{0.7}
\plotone{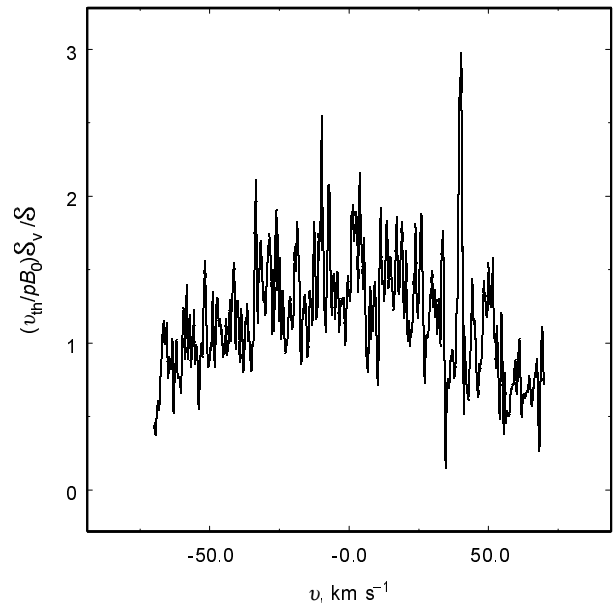}
\caption{%
The ratio from Figure 8 of the Stokes-$V$ flux ${\cal S}_V$ calculated
for a constant magnetic field $B_{\rm s}=B_0$ (bottom panel) to the energy
flux ${\cal S}$ (top panel) expressed as a dimensionless ratio.\label{fig10}}
\end{figure}

\end{document}